\documentclass[aps,twocolumn,showpacs,prd,nofootinbib,preprintnumbers]{revtex4}
\usepackage{graphicx}
\usepackage{dcolumn}
\usepackage{bm}
\usepackage{amsmath, amssymb}
\usepackage{color}

\begin{document}
\preprint{EPHOU-12-007}

\preprint{OU-HET 759/2012}

\title
{\Large\bf Stability of Leptonic Self-complementarity}

\author{Naoyuki Haba$^1$}
\author{Kunio Kaneta$^{1,2}$}
\author{Ryo Takahashi$^1$}

\affiliation{
$^{1}$Department of Physics, Faculty of Science, Hokkaido University, Sapporo 
060-0810, Japan}
\affiliation{$^{2}$Department of Physics, Graduate School of Science, Osaka 
University, Toyonaka, 560-0043, Japan}

\begin{abstract}
We investigate a stability of leptonic self-complementarity such that sum of 
three mixing angles in lepton sector is 90 degrees. Current experimental data of
 neutrino oscillation indicates that the self-complementarity can be satisfied 
within 3$\sigma$ ranges of each mixing angles. Thus the self-complementarity may
 be a key to study a flavor physics behind the standard model, and important to 
discuss its stability. We analyze renormalization group equations in a context 
of minimal supersymmetric standard model for the self-complementarity. It is 
seen that one of Majorana phases plays an important role for the stability of 
self-complementarity. We find some stable solutions against quantum corrections 
at a low energy. An effective neutrino mass for neutrino-less double beta decay 
is also evaluated by the use of neutrino parameters giving rise to the stable 
solutions.

\noindent
\end{abstract}
\maketitle

Neutrino oscillation experiments established that there are two large mixing 
angles ($\theta_{12}$ and $\theta_{23}$) of Pontecorvo-Maki-Nakagawa-Sakata (PMNS)
 matrix in lepton sector. Then a non-vanishing $\theta_{13}$ in the PMNS has been
 reported by recent long baseline and reactor neutrino 
experiments~\cite{An:2012eh}. These results can be interpreted by three flavor 
mixing of neutrinos. Regarding with neutrino masses $m_i$ ($i=1,2,3$), the 
neutrino oscillation experiments determine only two mass squared differences, 
$\Delta m_{21}^2\equiv|m_2|^2-|m_1|^2$ and $|\Delta 
m_{31}^2|\equiv||m_3|^2-|m_1|^2|$. Therefore, two types of neutrino mass hierarchy
 are allowed, i.e. normal hierarchy (NH) $m_1<m_2<m_3$ and inverted hierarchy 
(IH) $m_3<m_1<m_2$. Further, neutrino experiments have not determined whether 
the neutrinos are Dirac or Majorana particles. Clearly, the nature of neutrinos 
would be a key to find physics beyond the standard model (SM).

In theoretical side of neutrino physics, various approaches have been discussed 
in order to investigate hidden flavor structure behind the SM, e.g. 
introductions of flavor symmetry, mass (matrix) texture analyses, and searches 
for exotic relations among flavor mixing angles etc.. In this work, we focus 
on a leptonic self-complementarity~\cite{Zheng:2011uz} (see 
also~\cite{Luo:2012ce} for related discussions) as
 \begin{eqnarray}
  \theta_{12}+\theta_{23}+\theta_{13}=\frac{\pi}{2}=90^\circ. \label{90}
 \end{eqnarray}
The current experimental data of neutrino oscillation indicates that the 
self-complementarity can be satisfied within $3\sigma$ ranges of each mixing 
angles. Therefore, the self-complementarity may be a key to investigate a flavor
 physics behind the SM, and important to discuss its stability.\\

We start with effective Yukawa interaction and Weinberg operator at a low 
energy scale such as electroweak (EW) scale $\Lambda_{\rm EW}$ in a context of 
minimal supersymmetric standard model (MSSM), 
 \begin{equation}
  \mathcal{L}_Y=-y_e\overline{L_L}H_de_R+\frac{\kappa}{2}(H_uL_L)(H_uL_L)+h.c.,
 \end{equation} 
where $L_L$ are left-handed lepton doublets, $e_R$ are right-handed charged 
leptons, $H_u(H_d)$ is up(down)-type Higgs, $y_e$ is Yukawa matrix of charged 
leptons, and $\kappa(H_uL_L)(H_uL_L)$ is the Weinberg operator, which can be 
effectively induced by integrating out a heavy particle(s). One of examples to 
obtain this operator is seesaw mechanism. Typical scale of the seesaw mechanism 
is $\mathcal{O}(10^{14})$ GeV. Therefore, note that the effective coupling 
$\kappa$ is having mass dimension $-1$ and 
$\kappa^{-1}\sim\mathcal{O}(10^{14})$ GeV. Such a heavy mass scale can realize 
tiny active neutrino mass scales through the seesaw mechanism. In this work, we
 utilize an useful parameterization for the PMNS matrix 
as~\cite{Fritzsch:2001ty}
\begin{widetext}
 \begin{align}
  &V_{\rm PMNS}\equiv V_{eL}^\dagger V_\nu D_p 
  =\left(
               \begin{array}{ccc}
                c_{12}c_{13} & s_{12}c_{13} & s_{13} \\
                -c_{12}s_{23}s_{13}-s_{12}c_{23}e^{-i\delta} & -s_{12}s_{23}s_{13}+c_{12}c_{23}e^{-i\delta} & s_{23}c_{13} \\
                -c_{12}c_{23}s_{13}+s_{12}s_{23}e^{-i\delta} & -s_{12}c_{23}s_{13}-c_{12}s_{23}e^{-i\delta} & c_{23}c_{13} 
               \end{array}
              \right) 
              \left(
               \begin{array}{ccc}
                e^{i\rho} & 0 & 0 \\
                0 & e^{i\sigma} & 0 \\
                0 & 0 & 1 
               \end{array}
              \right),
 \end{align}
\end{widetext}
where $s_{ij}\equiv\sin\theta_{ij}$, $c\equiv\cos\theta_{ij}$ 
$(i,j=1,2,3;~i<j)$, $\delta$ is a Dirac phase, and $D_p$ is a diagonal phase 
matrix including two Majorana phases, $\rho$ and $\sigma$. An neutrino mass 
matrix $M_\nu$ can be diagonalized as $V_\nu^\dagger M_\nu 
V_\nu^\ast=M_\nu^{\rm diag}\equiv\mbox{Diag}\{m_1,m_2,m_3\}$ with 
$m_i\equiv\kappa_iv_u^2$ where $v_u$ is vacuum expectation value of up-type 
Higgs.

Next, we consider renormalization group equations (RGEs) in the MSSM. The RGE 
of $\kappa$ is given by 
$16\pi^2(d\kappa/dt)=\alpha\kappa+[(y_ey_e^\dagger)\kappa+\kappa(y_ey_e^\dagger)^T]$
with $\alpha\equiv6[-g_1^2/5-g_2^2+\mbox{Tr}(y_u^\dagger y_u)]$ where $g_i$ are
 gauge coupling constants, $t$ is an arbitrary renormalization scale as 
$t\equiv\ln(\mu/\Lambda)$, and $\Lambda$ is a high energy scale such as the 
seesaw scale~\cite{Chankowski:1993tx,Luo:2005sq}. One can also obtain RGEs of 
$\theta_{ij}$ in a diagonal basis of $y_e$ as 
$d\theta_{ij}/dt=F_{ij}(\theta_{12},\theta_{23},\theta_{13},\kappa_i,\delta,\rho,\sigma,y_\tau;t)$ where right-hand side (RHS) of this equation is given 
in~\cite{Luo:2005sq} (see also \cite{Casas:1999tg} for other discussions of 
mixing angles under the RGEs). Now we turn to the self-complementarity relation
 \eqref{90} and investigate the following equation,
 \begin{align}
  \frac{d}{dt}\sum_{ij}\theta_{ij}
  &=\sum_{ij}F_{ij}(\theta_{12},\theta_{23},\theta_{13},\kappa_i,
                             \delta,\rho,\sigma,y_\tau;t) \nonumber\\
  &\equiv F(\theta_{12},\theta_{23},\theta_{13},\kappa_i,\delta,\rho,\sigma,
            y_\tau;t), \label{sum}
 \end{align}
where $ij$ is summed over 12, 23, and 13. The function $F$ is described by 3 
mixing angles, 3 effective couplings for the light neutrino masses (or 
equivalently light neutrino masses $m_i$), 3 CP-phases, a Yukawa coupling of 
$\tau$, and renormalization scale. Then once we impose \eqref{90} on 
\eqref{sum} at an energy scale $t_0$, one of mixing angles in $F$ is removed as
 e.g. 
$\tilde{F}(\theta_{12},\theta_{23},\kappa_i,\delta,\rho,\sigma,y_\tau;t_0)$. We
 now focus on an equation,
 \begin{eqnarray}
  \tilde{F}(\theta_{12},\theta_{23},\kappa_i,\delta,\rho,\sigma,y_\tau;t_0)=0.
  \label{sol}
 \end{eqnarray}
This equation means that once the equation is satisfied at an energy scale 
$t_0$, the self-complementarity is also satisfied at all other energy scales 
$t$, i.e. the self-complementarity is stable against quantum corrections, if 
running effects of parameters except for mixing angles are tiny. In fact, we 
can find consistent solutions of \eqref{sol} with experiments for both NH and 
IH cases. According to the latest experimental data of neutrino 
oscillation~\cite{Tortola:2012te}
 \begin{align}
  &31.3^\circ\lesssim\theta_{12}
    \lesssim37.5^\circ, \label{ex12} \\ 
  &38.6^\circ\lesssim\theta_{23}
    \lesssim53.1^\circ, \label{ex23}\\
  &7.0^\circ(7.3^\circ)\lesssim\theta_{13}\lesssim
    10.9^\circ(11.1^\circ), \label{ex13}
 \end{align}
at 3$\sigma$ level for the NH(IH), the \eqref{90} can be satisfied. 

Mass spectra of neutrinos at a low energy are defined by 
$m_1\equiv\sqrt{m_3^2-|\Delta m_{31}^2|}$ and $m_2\equiv\sqrt{m_3^2-|\Delta 
m_{31}^2|+\Delta m_{21}^2}$ with best fit values 
$\Delta m_{21}^2=7.62\times10^{-5}\mbox{ eV}^2$ and $\Delta 
m_{31}^2=2.53\times10^{-3}\mbox{ eV}^2$ for the NH, and 
$m_1\equiv\sqrt{m_2^2-\Delta m_{21}^2}$ and $m_3\equiv\sqrt{m_2^2-|\Delta 
m_{31}^2|-\Delta m_{21}^2}$ with $\Delta m_{31}^2=2.40\times10^{-3}\mbox{ 
eV}^2$ for the IH. Therefore, the largest neutrino mass $m_3$($m_2$), of NH(IH)
 case is a free parameter in our analyses.  We analyze in range of 
$\sqrt{|\Delta m_{31}^2|}\leq m_3\leq0.2$ eV ($\sqrt{|\Delta m_{31}^2|+\Delta 
m_{21}^2}\leq m_2\leq0.2$ eV). The case of $m_3(m_2)=0.2$ eV corresponds to a 
degenerate mass spectrum. In such case, $m_3$($m_2$) is bounded by a 
cosmological constraint on the sum of neutrino mass as $\sum m_i\lesssim0.6$ 
eV~\cite{Komatsu:2010fb}, and thus as $m_3(m_2)\lesssim\sum m_i/3\simeq0.2$ eV.
 Therefore, this must be implied as upper bound on the largest neutrino mass. 
Here note that since input values of neutrino parameters at a low energy are 
used, the solutions of~\eqref{sol} are corresponding to ones at 
$t_0\simeq\ln(\Lambda_{\rm EW}/\Lambda)$. The $y_\tau$ has been also 
approximated at a low energy as $y_\tau(\Lambda_{\rm EW})=10^{-2}$ in our 
analyses.

There are 7 parameters (3 mixing angles, 3 CP-phases, and 1 neutrino mass, 
$m_3$ or $m_2$) and 2 imposed equations (\eqref{90} and \eqref{sol}). 
Therefore, number of free parameters is 5. Since it is however intricate to 
deal with all 5 parameters as completely free ones, we numerically analyze at 
some fixed neutrino masses as examples. According to our analyses, a CP-phase 
is important to give solutions of \eqref{sol} with \eqref{90}; we numerically 
found that there is no solution to satisfy \eqref{sol} in cases of 
($\delta=\rho=\sigma=0$) and ($\delta\neq0,~\rho=\sigma=0$) for both NH and IH,
 but ($\rho\neq0,~\delta=\sigma=0$) and ($\sigma\neq0,~\delta=\rho=0$) can give
 solutions of \eqref{sol} in some cases of NH. In a case of NH with a minimal 
$m_3$ (i.e. $m_3=\sqrt{|\Delta m_{31}^2|}$) and all cases of IH, neither 
($\sigma\neq0,~\delta=\rho=0$) nor ($\rho\neq0,~\delta=\sigma=0$) can give the 
solution. Therefore, in the following, we focus on the cases of 
($\rho\neq0,~\delta=\sigma=0$) and ($\sigma\neq0,~\delta=\rho=0$) for other 
cases of NH in detail. Now we have 2 free parameters (one of Majorana phases 
and one of mixing angles) in order to look for solution, i.e. once we fix one 
of Majorana phases and one of mixing angles, all values of our parameters are 
uniquely determined as we will explain below.

\begin{figure}
(a) $m_3=\sqrt{|\Delta m_{31}^2|+\Delta m_{21}^2}$ and 
($\rho\neq0,~\delta=\sigma=0$)

\includegraphics[scale=0.25]{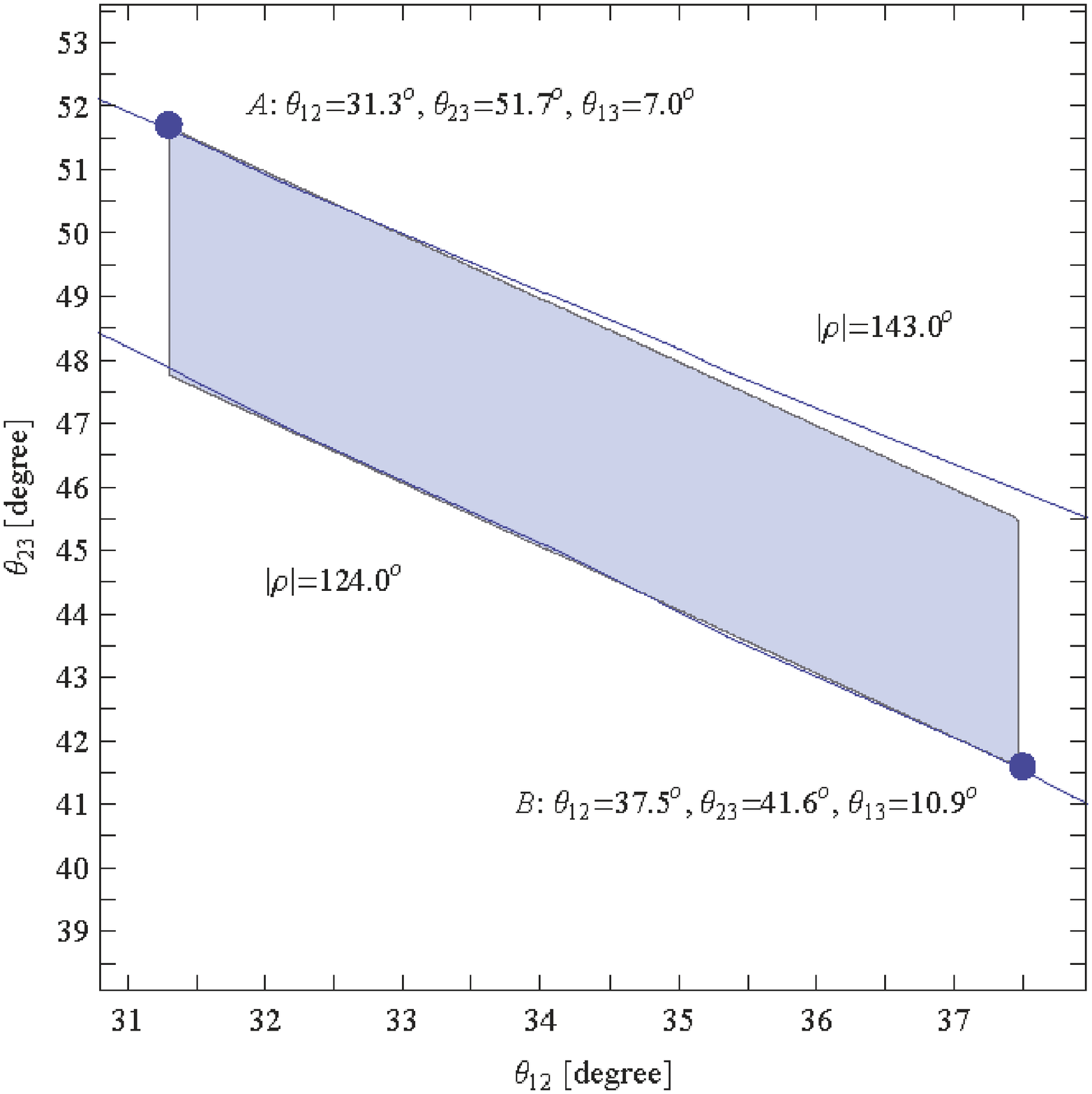}\vspace{3mm}

(b) $m_3=\sqrt{|\Delta m_{31}^2|+\Delta m_{21}^2}$ and 
($\sigma\neq0,~\rho=\sigma=0$)

\includegraphics[scale=0.25]{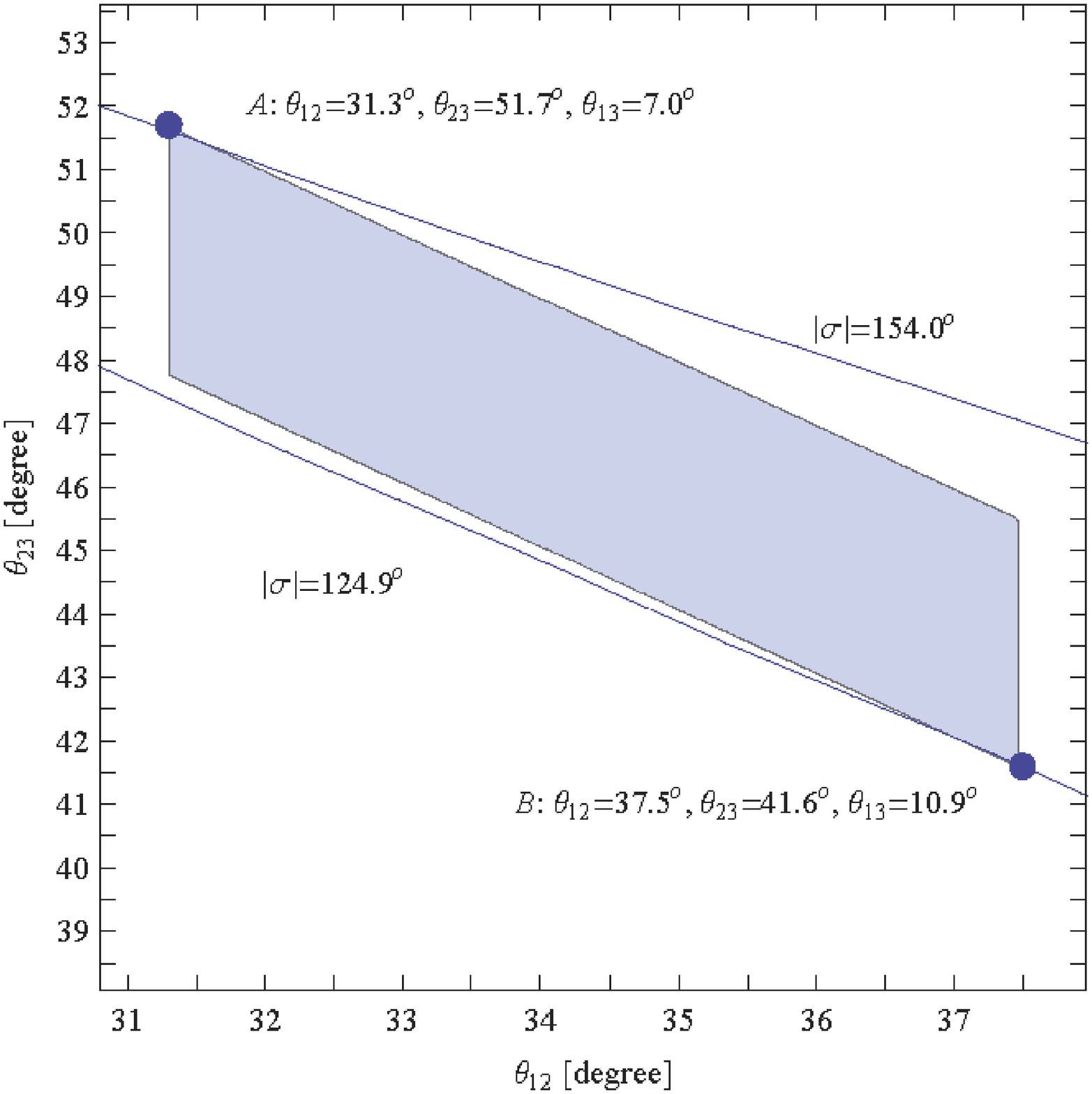}\vspace{3mm}

(c) $m_3=0.2$ eV and ($\sigma\neq0,~\rho=\sigma=0$)

\includegraphics[scale=0.25]{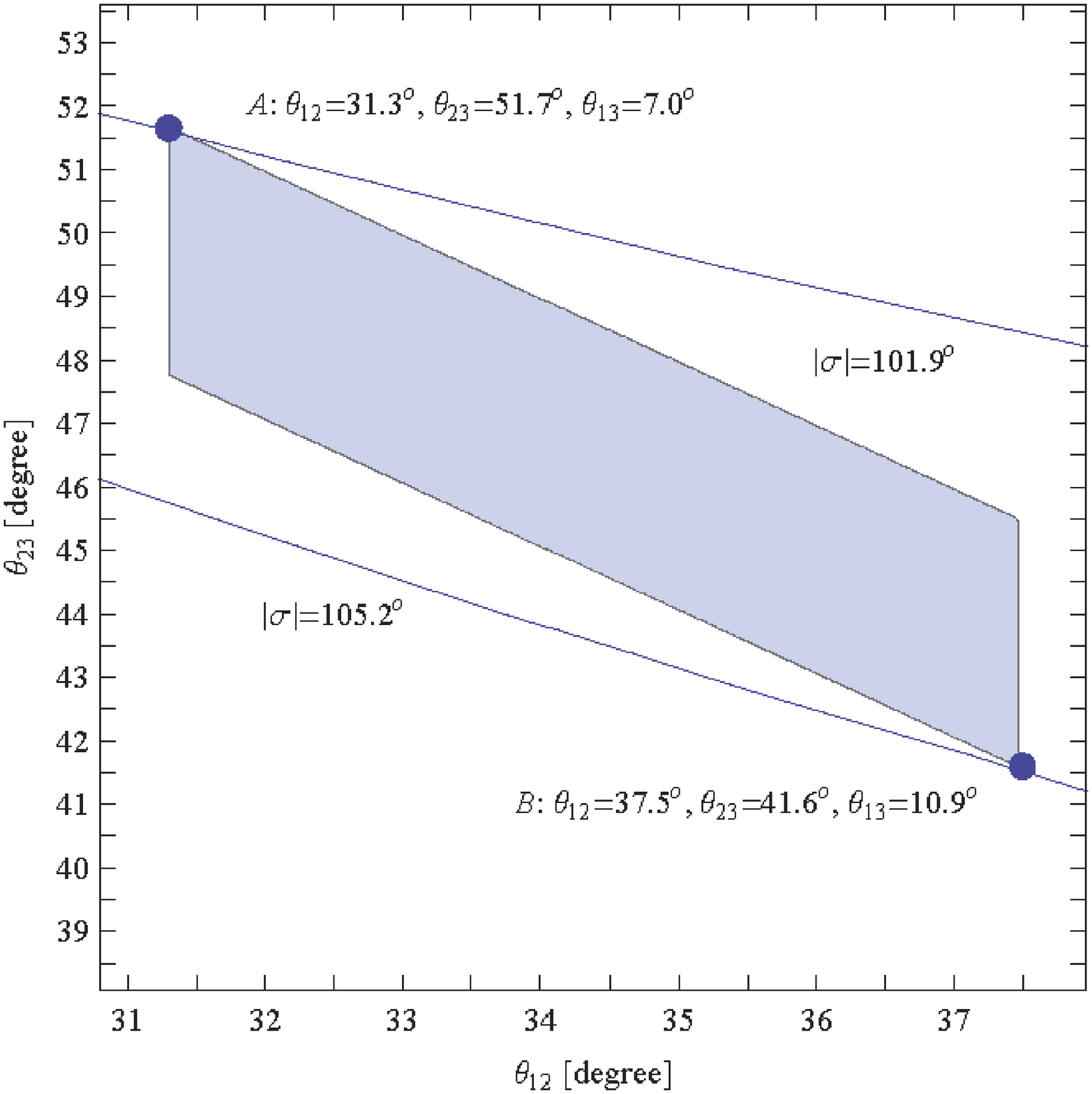}
\caption{Examples of solutions of \eqref{sol}}
\label{fig1}
\end{figure}

We have scanned over $0\leq(|\rho|\mbox{ or }|\sigma|)\leq\pi$. Some results of
 numerical analyses are shown in FIG.~\ref{fig1} as examples.
In the figures, vertical and horizontal axes are $\theta_{23}$ and 
$\theta_{12}$, respectively, and shaded region mean that the 
self-complementarity \eqref{90} is correlatively satisfied within $3\sigma$ 
ranges of mixing angles \eqref{ex12}-\eqref{ex13}. Lower(upper) slanting and 
right(left) sides are bounded by maximal(minimal) $\theta_{13}$ and 
$\theta_{12}$ at 3$\sigma$ level, respectively. Note that an allowed range of 
$\theta_{23}$ becomes narrow as 
$41.6^\circ\lesssim\theta_{23}\lesssim51.7^\circ$ compared to \eqref{ex23} due 
to \eqref{90}. Both two lines in the figures are solutions of \eqref{sol}, 
which are contours of $\rho$ or $\sigma$. FIG.~\ref{fig1} (a) is case of 
$m_3=\sqrt{|\Delta m_{31}^2|+\Delta m_{21}^2}$ with 
($\rho\neq0,~\delta=\sigma=0$) for the NH, and there are 2 remaining 
parameters, $\rho$ and one of mixing angles. In the case, 2 lines corresponding
 to $|\rho|\simeq124.0^\circ$ and $143.0^\circ$ are grazing shaded region at 
upper left described by A and lower right B points, respectively (there are 2 
(different sign) solutions of $\rho$ or $\sigma$ in all cases, i.e. solutions 
are symmetric for reflection respect with $\pi$). Therefore, values of all 7 
parameters in our analyses are uniquely determined at e.g. point A or B. 
Further, the points A and B in all cases determine maximal and minimal values 
of parameters for the stability of self-complementarity. Therefore, the 
self-complementarity is stable in shaded regions of 
$124.0^\circ\lesssim|\rho|\lesssim143.0^\circ$, 
$31.3^\circ\lesssim\theta_{12}\lesssim37.5^\circ$, 
$7.0^\circ\lesssim\theta_{13}\lesssim10.9^\circ$, and 
$\mbox{Min}[\theta_{23}]\leq\theta_{23}\leq\mbox{Max}[\theta_{23}]$ where 
$\mbox{Max(Min)}[\theta_{ij}]$ is maximal(minimal) value of $\theta_{ij}$. 
$\mbox{Max(Min)}[\theta_{23}]$ is evaluated by 
$\mbox{Max(Min)}[\theta_{23}]=90^\circ-\mbox{Min(Max)}[\theta_{12}]-\mbox{Min(Max)}[\theta_{13}]$ 
due to \eqref{90}. Note that value of one of mixing angle (e.g. $\theta_{23}$) 
is not independently taken because of \eqref{90}. In the case, we obtain 
$\mbox{Max(Min)}[\theta_{23}]\simeq51.7^\circ(41.6^\circ)$. The results are 
summarized in TAB.~\ref{tab1}.
\begin{table*}
\begin{center}
\begin{tabular}{c||c|c|c}
\hline
 & \multicolumn{3}{|c}{NH} \\
\hline\hline
$m_3$ & \multicolumn{2}{|c|}{$\sqrt{|\Delta m_{31}^2|+\Delta m_{21}^2}$} & 0.2 eV \\
\hline
$m_2$ & \multicolumn{2}{|c|}{$1.23\times10^{-2}$ eV} & 0.194 eV \\
\hline
$m_1$ & \multicolumn{2}{|c|}{$8.73\times10^{-3}$ eV} & 0.194 eV \\
\hline
Phases & $\rho\neq0,~\delta=\sigma=0$ & $\sigma\neq0,~\delta=\rho=0$ & $\sigma\neq0,~\delta=\rho=0$ \\
\hline
Min[$\theta_{12}$($\sin^2\theta_{12}$)] & \multicolumn{3}{|c}{$31.3^\circ(0.27)$} \\
\hline
Max[$\theta_{12}$($\sin^2\theta_{12}$)] & \multicolumn{3}{|c}{$37.5^\circ(0.37)$} \\
\hline
Min[$\theta_{23}$($\sin^2\theta_{23}$)] & \multicolumn{3}{|c}{$41.6^\circ(0.44)$} \\
\hline
Max[$\theta_{23}$($\sin^2\theta_{23}$)] & \multicolumn{3}{|c}{$51.7^\circ(0.62)$} \\
\hline
Min[$\theta_{13}$($\sin^2\theta_{13}$)] & \multicolumn{3}{|c}{$7.0^\circ(0.015)$} \\
\hline
Max[$\theta_{13}$($\sin^2\theta_{13}$)] & \multicolumn{3}{|c}{$10.9^\circ(0.036)$} \\
\hline
Min$[|\rho|\mbox{ or }|\sigma|]$ & $|\rho|=124.0^\circ$ & $|\sigma|=124.9^\circ$ & $|\sigma|=101.9^\circ$ \\
\hline
Max$[|\rho|\mbox{ or }|\sigma|]$ & $|\rho|=143.0^\circ$ & $|\sigma|=153.9^\circ(154.0^\circ)$ & $|\sigma|=104.6^\circ(105.2^\circ)$ \\
\hline\hline
Min$[\langle m\rangle_{ee}]$ [meV] & $6.08$ & $6.33$ & $97.3$ \\
\hline
Max$[\langle m\rangle_{ee}]$ [meV] & $9.13$ & $9.42$ & $72.7$ \\
\hline
\end{tabular}
\end{center}
\caption{Examples of solutions, and minimal and maximal values of neutrino 
parameters in the corresponding regions: Values of $|\sigma|$ in parentheses 
are maximal ones giving solutions of \eqref{sol} but the solutions realized by 
these maximal values are not stable against running effects of CP-phases. The 
values of $|\rho|$ and $|\sigma|$ without parentheses are complete stable ones 
against the running effects of CP-phases.}
\label{tab1}
\end{table*}
 
FIG.~\ref{fig1} (b) shows a case of $m_3=\sqrt{|\Delta m_{31}^2|+\Delta 
m_{21}^2}$ with ($\sigma\neq0,~\delta=\rho=0$). In the case, we obtain a region
 of $|\sigma|$ as $124.9^\circ\lesssim|\sigma|\leq154.0^\circ$, where the 
self-complementarity is satisfied, in a similar analysis to the previous case. 
Regions of mixing angles for the realization of the self-complementarity is the
 same as the previous case.

We have also analyzed a case of $m_3=0.2\mbox{ eV}$. Results are given in 
TAB.~\ref{tab1}. We cannot obtain any solutions of (\ref{sol}) in the case of 
($\rho\neq0,~\delta=\sigma=0$) but can do in one of 
($\sigma\neq0,~\delta=\rho=0$). In the case, values of $|\sigma|$ within 
$101.9^\circ\lesssim|\sigma|\lesssim105.2^\circ$ can give the solutions. 
Allowed region of mixing angles are the same as ones in the $m_3=\sqrt{|\Delta 
m_{31}^2|}$ case, i.e. contours of solutions can reach at the both points A and
 B like the case of $m_3=\sqrt{|\Delta m_{31}^2|}$ . For the 
Max(Min)$[\sigma]$, it is determined by Min(Max)$[\theta_{23}]$ in contrast 
with the above two cases.

Of course, other parameters (CP-phases and neutrino masses) contributing to 
mixing angles evolve under the RGEs. First, we comment on running effects of 
CP-phases. It has been seen that the Majorana phases are important for the 
stability of self-complementarity. One may worry about running effects of phases
 on low energy solutions for the stability, i.e. whether such effects spoil the 
solutions at a high energy scale or not. We have approximated the running 
effects of $\delta,~\rho$, and $\sigma$ from the seesaw scale on solutions by 
using leading-log estimation in the RGEs. These running effects 
$(\Delta\delta,\Delta\rho,\Delta\sigma)$ are 
$(\Delta\delta,\Delta\rho,\Delta\sigma)\sim(\mathcal{O}(0.1^\circ),\mathcal{O}(0.1^\circ),\mathcal{O}(0.1^\circ))$ 
and $(\mathcal{O}(1^\circ),\mathcal{O}(1^\circ),\mathcal{O}(1^\circ))$ for the 
cases of $m_3=\sqrt{|\Delta m_{31}^2|+\Delta m_{21}^2}$ and 0.2 eV, 
respectively. Therefore, most regions of Majorana phases are stable against 
such small running effects up to the seesaw scale because of 
Max$[\sigma]-\mbox{Min}[\sigma]>\Delta\sigma$. In fact, we show complete stable
 values of $|\rho|$ and $|\sigma|$, which are described by values without 
parentheses in TAB.~\ref{tab1}, including the above running effects from 
CP-phases. Further, small (but non-vanishing) running effect of $\delta$ does 
not affect the stability of self-complementarity. Then one must remember that 
our analyses are only for the Majorana neutrino case.

Next, we consider running effects of neutrino masses. We have also evaluated the
 effects from neutrino masses by the use of leading-log approximation in 
corresponding RGEs for the neutrino masses, which are also given 
in~\cite{Luo:2005sq}. These running effects from the seesaw scale to the 
electroweak one are 
 \begin{align}
  (\Delta m_1^{eff}&,\Delta m_2^{eff},\Delta m_3^{eff}) \nonumber \\
  &\sim
   \left\{
    \begin{array}{l}
     (\mathcal{O}(10^{-8}),\mathcal{O}(10^{-7}),\mathcal{O}(10^{-6})) \mbox{ eV} \\ 
     (\mathcal{O}(10^{-6}),\mathcal{O}(10^{-6}),\mathcal{O}(10^{-6})) \mbox{ eV}
    \end{array}
   \right.,
 \end{align} 
for the cases of $m_3=\sqrt{|\Delta m_{31}^2|+\Delta m_{21}^2}$ and 0.2 eV, 
respectively, where $\Delta m_i^{eff}$ affects only evolutions of mixing angles
 and CP-phases not absolute values of neutrino mass eigenvalues, i.e. 
overall (flavor mixing independent) contributions from running of Yukawa 
couplings (top Yukawa gives dominant contribution) are omitted. Even with these
 running effects of neutrino masses, the solutions given in the TAB.~\ref{tab1}
 are stable, i.e. we can also obtain solutions of \eqref{sol} within the almost
 same range of $|\sigma|$ as ones in TAB.~\ref{tab1}. We have also numerically 
checked evolutions of mixing angles and their sum in order to make sure that 
the one of solutions in the NH is stable. In the calculation, we take
\begin{align}
& m_1=8.73\times10^{-3}\mbox{ eV}, \\
& m_2=1.23\times10^{-2}\mbox{ eV}, \\
& m_3=\sqrt{|\Delta m_{31}^2|+\Delta m_{21}^2}, \\
& \delta=\sigma=0,~\rho=133.5^\circ,
\end{align}
at low energy as an example and
\begin{eqnarray}
\theta_{12}=33.5^\circ,~\theta_{23}=47.7^\circ,~\theta_{13}=8.8^\circ,
\end{eqnarray}
as low energy boundary conditions for the RGEs. The running effects of the 
mixing angles from the seesaw scale to the electroweak one are
 \begin{align}
  &(\Delta\theta_{12},\Delta\theta_{23},\Delta\theta_{13})\sim \nonumber \\
  &(-\mathcal{O}(10^{-4}),\mathcal{O}(10^{-3}),-\mathcal{O}(10^{-4}))
   \mbox{ [degree]}.
 \end{align}
Since the deviation of the leptonic self-complementarity from $90^\circ$ is 
$\mathcal{O}(10^{-4})$ degree, the self-complementarity relation can be still 
stable.

Finally, we evaluate effective mass term of neutrino-less double beta decay 
($0\nu\beta\beta$), $\langle m\rangle_{ee}\equiv|\sum_{i=1}^3(V_{\rm 
PMNS})_{ei}^2m_i|$, in our parameter space. It is written down as 
 \begin{align}
  \langle m\rangle_{ee} 
   &=\left|m_1c_{12}^2c_{13}^2e^{2i\rho}+m_2s_{12}^2c_{13}^2e^{2i\sigma}+m_3s_{13}^2\right|, \label{mee}
 \end{align}
in our notation. The phenomenon of $0\nu\beta\beta$ can distinguish whether 
neutrinos are Dirac or Majorana particles. The results at benchmarks given in 
Tab.~\ref{tab1} are also presented in the table.

The magnitude of $\langle m\rangle_{ee}$ strongly depends on the scale of $m_1$
 or $m_2$ rather than mixing angles and CP-phases in the cases. In the NH with 
$m_3=\sqrt{|\Delta m_{31}^2|}$ and $\sqrt{|\Delta m_{31}^2|+\Delta m_{21}^2}$, 
dominant contribution comes from the second term of RHS of \eqref{mee} because 
of the small $s_{13}^2$ and vanishing $m_1$. We predict 6.08 
meV$\lesssim\langle m\rangle_{ee}\lesssim97.3$ meV for the NH within the 
parameter space to make the self-complementarity stable. The Heidelberg-Moscow 
experiment~\cite{KlapdorKleingrothaus:2000sn} for $0\nu\beta\beta$ is giving 
the most severe bound on $\langle m\rangle_{ee}$, which is $\langle 
m\rangle_{ee}\lesssim 210$ meV. The CUORE experiment~\cite{Arnaboldi:2002du} is
 expected to reach $\langle m\rangle_{ee}=(24-93)$ meV in the future. 
Therefore, a part of our predicting region may be checked in future 
experiments.\\

We have investigated a stability of leptonic self-complementarity relation in 
the PMNS sector against quantum corrections by considering RGEs in the  MSSM. 
The current experimental data of neutrino oscillation indicates that that the 
self-complementarity can be satisfied at 3$\sigma$ ranges of each mixing angle. 
This motivates us to study the self-complementarity and its stability as a key 
to find a physics behind the SM. As the results of analyses, we have found 
solutions stabilizing the self-complementarity by using low energy data of 
neutrino oscillation experiments. It has been seen that the Majorana play an 
important role to give the solutions. The self-complementarity relation can be 
satisfied up to an arbitrary high energy scale if neutrino parameters are 
correlatively within $31.3^\circ\lesssim\theta_{12}\lesssim37.5^\circ$, 
$7.0^\circ\lesssim\theta_{13}\lesssim10.9^\circ$, and 
Max(Min)$[\theta_{23}]\simeq51.7^\circ(41.6^\circ)$
with $124.0^\circ\lesssim|\rho|\lesssim143.0^\circ$ or 
$124.9^\circ\lesssim|\sigma|\lesssim153.9^\circ$ for $m_3=\sqrt{|\Delta 
m_{31}^2|+\Delta m_{21}^2}$, and 
$101.9^\circ\lesssim|\sigma|\lesssim104.6^\circ$ for $m_3=0.2$ eV of NH at a 
low energy. These solutions and leptonic self-complementarity relation are 
stable against running effects of CP-phases and neutrino masses. Regarding with
 the $0\nu\beta\beta$, the effective neutrino mass can be predicted as $6.08$ 
meV$\lesssim\langle m\rangle_{ee}\lesssim 97.3$ meV for the stable solution in 
NH case of $m_3=\sqrt{|\Delta m_{31}^2|}$.\\ 

{\bf Acknowledgment:} 
This work is partially supported by Scientific Grant by 
Ministry of  Education and Science, Nos. 22011005, 24540272, 20244028, and 
21244036. The works of K.K. and R.T. are supported by Research Fellowships of 
the Japan Society for the Promotion of Science for Young Scientists.


\end{document}